\documentclass[11pt]{article}
\usepackage[final]{acl}
\usepackage{times}
\usepackage{latexsym}
\usepackage[T1]{fontenc}
\usepackage[utf8]{inputenc}
\usepackage{microtype}
\usepackage{inconsolata}
\usepackage{graphicx}
\usepackage{booktabs}
\usepackage{amsmath}
\usepackage{amssymb}
\usepackage{multirow}
\usepackage{xcolor}
\usepackage{pgfplots}
\pgfplotsset{compat=1.18}
\usepgfplotslibrary{fillbetween}
\usepackage{tikz}
\usetikzlibrary{shapes, arrows.meta, positioning, fit, backgrounds, calc, patterns}

\definecolor{vanillacolor}{HTML}{888888}
\definecolor{hybridcolor}{HTML}{4682B4}
\definecolor{accentred}{HTML}{C0392B}
\definecolor{accentgreen}{HTML}{27AE60}
\definecolor{lightblue}{HTML}{D6EAF8}
\definecolor{lightyellow}{HTML}{FEF9E7}
\definecolor{lightgreen}{HTML}{D5F5E3}
\definecolor{boxbg}{HTML}{F4F6F7}
\usepackage{fancyhdr}

\title{One Retrieval to Cover Them All: \\ Co-occurrence-Aware Knowledge Base Reorganization \\ for Session-Level RAG\thanks{Accepted to the Towards Knowledgeable Foundation Models (KnowFM) Workshop @ ACL 2026.}}

\author{Shivam Ratnakar \quad Yixuan Zhu \quad Cecilia Cheng \quad Chaya Vijayakumar \\
  University of Southern California \\
  \texttt{\{sratnaka, yzhu1458, huixiche, cvijayak\}@usc.edu}
}

\begin{document}
\maketitle

\pagestyle{fancy}
\fancyhf{}
\cfoot{\thepage}
\renewcommand{\headrulewidth}{0pt}
\begin{abstract}
RAG systems retrieve documents optimized for answering \emph{one query at a time}. Yet enterprise users arrive with \emph{sessions}, that is, coherent episodes of related questions that span semantically distant parts of the knowledge base. We show that a single retrieval call over a standard knowledge base covers only 41\% of a user's session-level information need. To close this gap, we reorganize the KB offline using co-occurrence-aware clustering and expand retrieval candidates through cluster neighborhoods at query time. On WixQA (6,221 enterprise support articles), our method raises single-query session coverage to 58\% (+17\% absolute; 95\% CI: [14.1, 20.4]), reduces retrieval calls to 70\% coverage by 34\%, and compresses the KB to 20\% of its original size, all consistently across four embedding models and six functional domains. We argue that session-level coverage, not single-query recall, should be the primary metric for enterprise RAG evaluation.
\end{abstract}

\section{Introduction}

Consider a user who contacts Wix customer support and asks: \emph{``How do I connect my own domain?''} A standard RAG system~\citep{lewis2021retrievalaugmented} retrieves domain-configuration articles and generates a helpful answer. But the user's actual need is broader: they also want to change their site template, configure payment processing, and set up email forwarding. These articles live in entirely different semantic neighborhoods of the knowledge base. The user must ask four separate questions, trigger four separate retrieval calls, and hope the system surfaces the right documents each time.

This scenario exposes a blind spot in how we build and evaluate RAG systems. The entire pipeline, from embedding models to retrieval metrics, is optimized for \emph{single-query relevance}: does the top-$k$ set contain the one document that answers this one question? But enterprise users do not arrive with isolated questions. They arrive with \emph{sessions}: coherent episodes of related information needs that span multiple documents across the knowledge base.

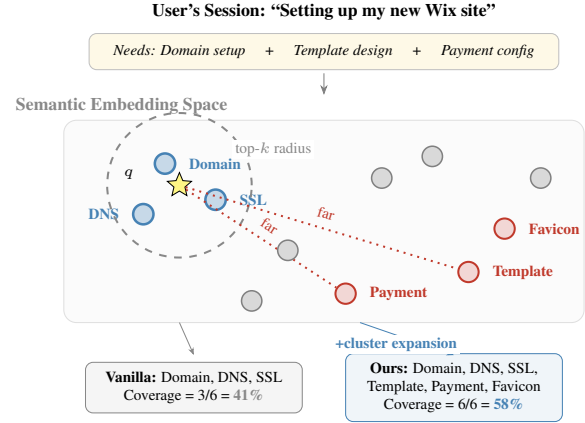
\begin{figure}[t]
\centering
\resizebox{\columnwidth}{!}{%
\begin{tikzpicture}[
    doc/.style={circle, minimum size=8pt, inner sep=0pt, draw=#1, fill=#1!30, line width=0.6pt},
    doc/.default=gray,
    qdoc/.style={circle, minimum size=8pt, inner sep=0pt, draw=hybridcolor, fill=hybridcolor!30, line width=1pt},
    rdoc/.style={circle, minimum size=8pt, inner sep=0pt, draw=accentred, fill=accentred!20, line width=0.8pt},
    lbl/.style={font=\tiny, #1},
    lbl/.default=black,
]

\node[font=\scriptsize\bfseries] at (0, 5.8) {User's Session: ``Setting up my new Wix site''};
\node[draw=gray!50, fill=lightyellow, rounded corners=3pt, font=\tiny\itshape,
      minimum width=6.5cm, minimum height=0.5cm] at (0, 5.25) {
  Needs: Domain setup \quad + \quad Template design \quad + \quad Payment config
};

\draw[-{Stealth[length=4pt]}, gray] (0, 4.95) -- (0, 4.65);

\node[font=\scriptsize\bfseries, vanillacolor] at (-2.8, 4.5) {Semantic Embedding Space};

\draw[gray!25, fill=gray!3, rounded corners=5pt] (-3.6, 1.5) rectangle (3.6, 4.3);

\node[qdoc] (d1) at (-2.2, 3.7) {};
\node[lbl=hybridcolor, right] at (-2.0, 3.7) {\textbf{Domain}};
\node[qdoc] (d2) at (-2.5, 3.0) {};
\node[lbl=hybridcolor, left] at (-2.7, 3.0) {\textbf{DNS}};
\node[qdoc] (d3) at (-1.5, 3.2) {};
\node[lbl=hybridcolor, right] at (-1.3, 3.2) {\textbf{SSL}};

\node[rdoc] (t1) at (2.0, 2.2) {};
\node[lbl=accentred, right] at (2.2, 2.2) {\textbf{Template}};
\node[rdoc] (t2) at (2.5, 2.8) {};
\node[lbl=accentred, right] at (2.7, 2.8) {\textbf{Favicon}};

\node[rdoc] (p1) at (0.3, 1.9) {};
\node[lbl=accentred, right] at (0.5, 1.9) {\textbf{Payment}};

\node[star, star points=5, star point ratio=2.2, draw=black, fill=yellow!60,
      minimum size=10pt, inner sep=0pt] (qpt) at (-2.0, 3.4) {};
\node[lbl, left] at (-2.5, 3.55) {$q$};

\draw[vanillacolor, thick, dashed] (-2.0, 3.4) circle (1.0cm);
\node[font=\tiny, vanillacolor, fill=white, inner sep=1pt] at (-0.7, 3.9) {top-$k$ radius};

\draw[accentred, dotted, thick] (qpt) -- (t1) node[midway, above, font=\tiny, accentred, sloped] {far};
\draw[accentred, dotted, thick] (qpt) -- (p1) node[midway, above, font=\tiny, accentred, sloped] {far};

\node[doc=gray] at (-0.5, 2.5) {};
\node[doc=gray] at (0.8, 3.5) {};
\node[doc=gray] at (1.5, 3.8) {};
\node[doc=gray] at (-1.0, 1.8) {};
\node[doc=gray] at (3.0, 3.5) {};


\node[draw=vanillacolor, fill=vanillacolor!8, rounded corners=3pt,
      minimum width=3cm, minimum height=0.7cm, font=\tiny, align=center] (vr) at (-1.8, 0.6) {
  \textbf{Vanilla:} Domain, DNS, SSL\\
  Coverage = 3/6 = \textcolor{vanillacolor}{\textbf{41\%}}
};

\node[draw=hybridcolor, fill=hybridcolor!8, rounded corners=3pt,
      minimum width=3cm, minimum height=0.7cm, font=\tiny, align=center] (hr) at (1.8, 0.6) {
  \textbf{Ours:} Domain, DNS, SSL,\\
  Template, Payment, Favicon\\
  Coverage = 6/6 = \textcolor{hybridcolor}{\textbf{58\%}}
};

\draw[-{Stealth[length=3pt]}, vanillacolor] (-2.0, 1.5) -- (-1.8, 1.05);
\draw[-{Stealth[length=3pt]}, hybridcolor] (0.5, 1.5) -- (1.8, 1.05);
\node[font=\tiny\bfseries, hybridcolor, fill=white, inner sep=1pt] at (1.0, 1.2) {+cluster expansion};

\end{tikzpicture}%
}
\caption{\textbf{The session coverage gap.} A user setting up a website needs articles about domains, templates, and payments, but these reside in distant regions of the semantic embedding space. Vanilla RAG retrieves only documents within the top-$k$ similarity radius of the query (dashed circle), missing semantically distant but session-relevant articles. Our cluster expansion recovers these, raising session coverage from 41\% to 58\%.}
\label{fig:hook}
\end{figure}

The core problem is that \textbf{semantic similarity and user information need are different signals}. Documents that a user needs together during a session do not necessarily look alike in embedding space. Domain configuration articles and template design articles serve the same user journey but reside in distant semantic neighborhoods. Standard RAG, which retrieves by embedding similarity alone, cannot bridge this gap. It finds documents that \emph{look like} the query but misses documents the user \emph{also needs}.

We formalize this intuition by introducing \textbf{session-level evaluation metrics} for RAG (Section~\ref{sec:metrics}). Single-query session coverage measures what fraction of a user's full information need is satisfied by one retrieval call. On WixQA~\citep{cohen2025wixqa}, we find that vanilla RAG covers only 41\% of the session at $k{=}8$, meaning the user must issue multiple follow-up queries to access the remaining knowledge.

To close this gap, we propose \textbf{co-occurrence-aware KB reorganization} (Section~\ref{sec:method}). The idea is simple: documents that users need together should be retrievable together. We learn a co-occurrence embedding space using Word2Vec~\citep{mikolov2013efficient} trained on document navigation sequences, cluster the KB in this space, and expand retrieval candidates through cluster neighborhoods at query time. This reorganization happens once, offline, and the resulting cluster structure benefits any downstream embedding model. While we primarily use synthetic co-occurrence sequences in this work, we note that the ground-truth article groups from WixQA's expert-labeled queries provide a real co-occurrence signal, and our method upsamples these 10$\times$ during training to anchor the learned representations in genuine user needs.

This approach has three appealing properties. First, it is \textbf{pre-retrieval}: the knowledge base is reorganized once offline, contrasting with post-retrieval methods like EDC$^2$-RAG~\citep{li2025efficient} that cluster after retrieval. Second, it is \textbf{encoder-agnostic}: because clusters are learned from co-occurrence patterns rather than from a specific embedding model, the same cluster structure improves retrieval across different encoders. Third, it provides \textbf{KB compression}: the cluster structure reduces the effective knowledge base to 20\% of its original size while improving coverage.

Our contributions are: (1) we introduce session-level evaluation metrics for RAG that capture multi-document information needs; (2) we propose co-occurrence-aware KB reorganization with cluster-expanded hybrid retrieval; and (3) we demonstrate consistent gains across four encoders, six domains, and four complexity levels on an enterprise benchmark.

\section{Related Work}

\paragraph{Query-Side RAG Optimization.} A large body of work optimizes RAG retrieval at query time. Multi-query approaches rewrite the user's query into multiple variants to improve recall. Adaptive retrieval methods like CAR~\citep{xu2025clusterbased} dynamically adjust how many documents to retrieve based on query complexity. HyDE~\citep{gao2023precise} generates hypothetical answers to improve query embeddings. Hybrid retrieval combines dense and sparse signals~\citep{ma2023finetuning}. All of these optimize \emph{which query} is sent to the retriever. Our work is complementary: we optimize \emph{how the knowledge base is organized} before any query arrives. The two approaches can be combined; for instance, multi-query retrieval could be applied on top of our reorganized KB, with each rewritten query benefiting from cluster expansion. We focus our baselines on the retrieval layer specifically because our contribution is pre-retrieval KB reorganization, not query rewriting.

\paragraph{Post-Retrieval Document Compression.} EDC$^2$-RAG~\citep{li2025efficient} clusters retrieved documents at query time to remove noise and redundancy before passing context to the LLM, and was accepted at EMNLP 2025 Findings. CRAG~\citep{akesson2024crag} similarly clusters and summarizes retrieved documents to fit context windows. RAPTOR~\citep{sarthi2024raptor} builds hierarchical summaries for tree-based retrieval. These methods operate \emph{after} retrieval to compress what was already fetched. Our method operates \emph{before} retrieval to ensure the right documents are fetched in the first place. The two approaches compose naturally: one can apply post-retrieval compression on top of our pre-retrieval reorganization.

\paragraph{Embedding Limitations.} \citet{weller2026theoretical} establish theoretical limits on single-vector dense embeddings, showing that fixed-dimensional representations cannot realize all possible retrieval configurations. This underscores the need for complementary signals beyond semantic similarity. Our co-occurrence embeddings provide exactly such a signal, capturing functional relatedness between documents that may be semantically distant.

\paragraph{Collaborative Filtering in Information Retrieval.} The use of co-occurrence patterns to learn item representations has a long history in recommendation systems. Item2Vec~\citep{barkan2017item2vec} applies Word2Vec to item co-purchase sequences. Our work applies the same principle to document co-access patterns in knowledge bases, bridging collaborative filtering and information retrieval for RAG.

\paragraph{Session-Level Evaluation.} Existing RAG evaluation focuses exclusively on single-query metrics such as precision@$k$, recall@$k$, MRR, and faithfulness~\citep{es2024ragas}. While some observability platforms support session-level monitoring for multi-turn conversations, no prior work evaluates retrieval \emph{coverage} at the session level or measures how many retrieval calls are needed to satisfy a user's full information need. We introduce these metrics and argue that they better reflect enterprise RAG performance, where user satisfaction depends on resolving an entire issue rather than answering a single question.

\section{Problem Formulation}
\label{sec:metrics}

Let $\mathcal{D} = \{d_1, \ldots, d_N\}$ be a knowledge base of $N$ documents. A \textbf{session} $\mathcal{S} \subseteq \mathcal{D}$ is a set of documents that collectively address a user's information-seeking episode. Given a query $q$ associated with session $\mathcal{S}$, a retriever returns an ordered set $\mathcal{R}_q = \{r_1, \ldots, r_k\}$ of $k$ documents.

\subsection{Session-Level Metrics}

We propose two metrics that evaluate retrieval at the session level rather than the query level.

\paragraph{Single-Query Session Coverage (Cov).} The fraction of the session's documents retrieved in one call:
\begin{equation}
    \text{Cov}(q, \mathcal{S}) = \frac{|\mathcal{R}_q \cap \mathcal{S}|}{|\mathcal{S}|}
\label{eq:coverage}
\end{equation}
Standard recall@$k$ is the special case where $|\mathcal{S}| = 1$. When $|\mathcal{S}| > 1$, Coverage captures multi-document information needs that recall@$k$ cannot distinguish. A Coverage of 0.6 means 60\% of the user's total information need is satisfied without any follow-up queries.

\paragraph{Calls to $\tau$-Coverage ($C_\tau$).} The minimum number of retrieval calls such that the union of all retrieved sets covers fraction $\tau$ of the session:
\begin{equation}
    C_\tau(\mathcal{S}) = \min \left\{ m : \frac{\left|\bigcup_{i=1}^{m} \mathcal{R}_{q_i} \cap \mathcal{S}\right|}{|\mathcal{S}|} \geq \tau \right\}
\label{eq:calls}
\end{equation}
where $q_1, \ldots, q_m$ are sequential queries from the session. Lower $C_\tau$ indicates a more efficient retrieval system. In production, each additional retrieval call adds latency, API cost, and user friction.

\section{Method}
\label{sec:method}

\begin{figure*}[t]
\centering
\begin{tikzpicture}[
    node distance=0.5cm,
    box/.style={rectangle, draw=#1!70, fill=#1!10, minimum width=1.7cm, minimum height=0.65cm, font=\scriptsize, rounded corners=3pt, line width=0.7pt, align=center},
    box/.default=gray,
    arr/.style={-{Stealth[length=4pt]}, thick, #1},
    arr/.default=gray!50,
    doc/.style={circle, draw=#1, fill=#1!25, minimum size=5pt, inner sep=0pt},
    doc/.default=gray,
]

\node[font=\small\bfseries] at (-4.5, 4.2) {(a) Offline: KB Reorganization};

\node[font=\scriptsize\bfseries, vanillacolor] at (-6.8, 3.6) {Semantic Space};
\draw[gray!30, rounded corners=3pt] (-8.2, 0.6) rectangle (-5.4, 3.3);

\node[doc=accentred] at (-7.8, 2.8) {};
\node[doc=accentred] at (-6.2, 1.2) {};
\node[doc=accentred] at (-7.0, 2.0) {};
\node[doc=hybridcolor] at (-7.5, 1.0) {};
\node[doc=hybridcolor] at (-5.8, 2.6) {};
\node[doc=hybridcolor] at (-6.5, 3.0) {};
\node[doc=accentgreen] at (-6.0, 1.8) {};
\node[doc=accentgreen] at (-7.6, 2.2) {};
\node[doc=accentgreen] at (-5.7, 0.9) {};
\node[doc=gray] at (-7.2, 1.4) {};
\node[doc=gray] at (-6.8, 0.8) {};
\node[doc=gray] at (-5.9, 2.2) {};

\node[font=\tiny, gray, text width=2.5cm, align=center] at (-6.8, 0.3) {Same-session docs are\\scattered across space};

\draw[arr=hybridcolor, very thick, -{Stealth[length=6pt]}] (-5.1, 2.0) -- (-3.7, 2.0);
\node[font=\tiny\bfseries, hybridcolor, text width=1.2cm, align=center] at (-4.4, 2.5) {Word2Vec\\on co-occ};

\node[font=\scriptsize\bfseries, hybridcolor] at (-2.0, 3.6) {Co-occurrence Space};
\draw[gray!30, rounded corners=3pt] (-3.4, 0.6) rectangle (-0.6, 3.3);

\draw[accentred!30, fill=accentred!8, rounded corners=3pt] (-3.2, 2.3) rectangle (-2.2, 3.1);
\node[doc=accentred] at (-2.9, 2.8) {};
\node[doc=accentred] at (-2.5, 2.6) {};
\node[doc=accentred] at (-2.7, 2.5) {};
\draw[hybridcolor!30, fill=hybridcolor!8, rounded corners=3pt] (-3.2, 0.8) rectangle (-2.2, 1.6);
\node[doc=hybridcolor] at (-2.9, 1.3) {};
\node[doc=hybridcolor] at (-2.5, 1.1) {};
\node[doc=hybridcolor] at (-2.7, 1.4) {};
\draw[accentgreen!30, fill=accentgreen!8, rounded corners=3pt] (-1.8, 1.5) rectangle (-0.8, 2.3);
\node[doc=accentgreen] at (-1.5, 2.0) {};
\node[doc=accentgreen] at (-1.1, 1.8) {};
\node[doc=accentgreen] at (-1.3, 2.1) {};

\node[font=\tiny, hybridcolor, text width=2.5cm, align=center] at (-2.0, 0.3) {Same-session docs are\\clustered together};

\node[font=\small\bfseries] at (4.2, 4.2) {(b) Online: Hybrid Retrieval};

\node[box=gray, minimum width=2cm] (q) at (2.0, 3.3) {Query $q$};

\node[box=gray] (enc) at (2.0, 2.3) {Encode\\(any model)};
\draw[arr] (q) -- (enc);

\node[box=vanillacolor, minimum width=2cm] (van) at (2.0, 1.2) {Top-$k_v{=}3$\\Vanilla Retrieval};
\draw[arr=vanillacolor] (enc) -- (van);

\node[doc=vanillacolor, minimum size=7pt] (d1) at (4.2, 1.7) {};
\node[doc=vanillacolor, minimum size=7pt] (d2) at (4.7, 1.2) {};
\node[doc=vanillacolor, minimum size=7pt] (d3) at (4.2, 0.7) {};
\draw[arr=vanillacolor] (van) -- (d1);
\node[font=\tiny, right] at (4.9, 1.2) {3 docs};

\draw[hybridcolor, very thick, dashed, rounded corners=5pt] (3.7, 0.2) rectangle (6.5, 2.2);
\node[font=\tiny\bfseries, hybridcolor, fill=white, inner sep=1pt] at (5.1, 2.2) {Cluster Expansion};

\node[doc=hybridcolor, minimum size=7pt] (e1) at (5.5, 1.6) {};
\node[doc=hybridcolor, minimum size=7pt] (e2) at (5.8, 1.1) {};
\node[doc=hybridcolor, minimum size=7pt] (e3) at (5.3, 0.6) {};
\node[doc=hybridcolor, minimum size=7pt] (e4) at (6.1, 0.7) {};
\node[doc=hybridcolor, minimum size=7pt] (e5) at (6.0, 1.5) {};
\node[font=\tiny, right] at (6.2, 1.1) {+12};

\draw[arr=accentred, very thick] (5.1, 0.0) -- (5.1, -0.7);
\node[font=\tiny\bfseries, accentred, right] at (5.2, -0.35) {Re-rank};

\node[box=accentgreen, minimum width=3cm] (res) at (5.1, -1.2) {Top-$k{=}8$ Results\\Coverage: 58\%};

\draw[arr=accentgreen, dashed, thick] (-0.6, 1.9) .. controls (1.5, 0.5) and (3.0, 0.0) .. (3.7, 1.2);
\node[font=\tiny, accentgreen, fill=white, inner sep=1pt] at (1.5, 0.2) {cluster structure};

\end{tikzpicture}
\caption{\textbf{System overview.} \textbf{(a) Offline:} Documents from the same user session (same color) are scattered in semantic embedding space but cluster together after Word2Vec training on co-occurrence sequences. \textbf{(b) Online:} Given a query, we retrieve the top-3 by similarity (gray dots), expand through their cluster neighborhoods (blue dots, +12 candidates), and re-rank the union to return the top-8 with 58\% session coverage.}
\label{fig:architecture}
\end{figure*}
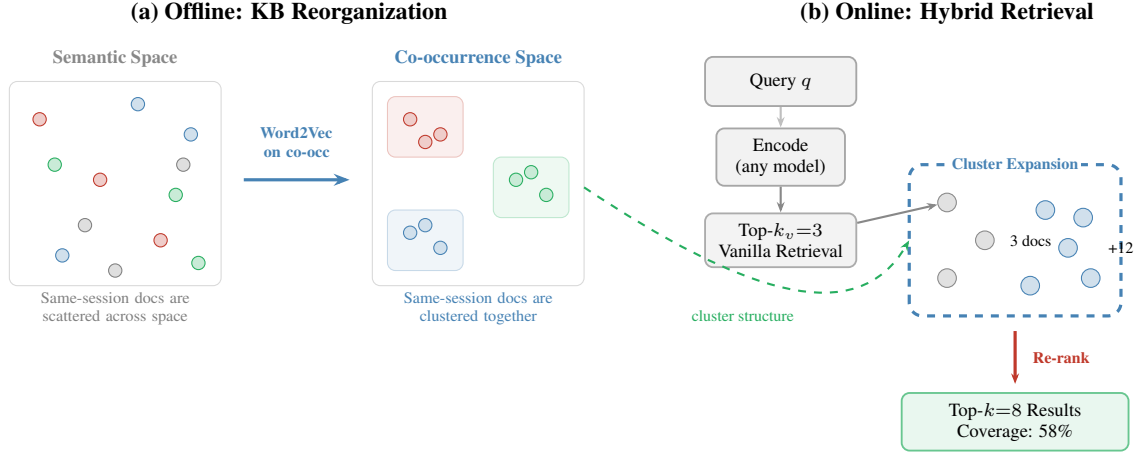

Our approach consists of an offline KB reorganization phase and an online hybrid retrieval phase (Figure~\ref{fig:architecture}).

\subsection{Offline: Co-occurrence-Aware Clustering}

\paragraph{Step 1: Co-occurrence Sequence Construction.} We construct sequences of document IDs that represent documents a user would need together. These sequences are drawn from three complementary sources:

\begin{enumerate}
    \item \textbf{Ground-truth co-occurrence:} For each labeled query with multiple relevant article IDs, the article set forms a natural co-occurrence group. These are the highest-quality signal, as they reflect real user needs annotated by domain experts.
    \item \textbf{Embedding-neighborhood walks:} Starting from a random seed document, we perform random walks through the document similarity graph for 3 to 5 steps, with a 40\% random jump probability. The high jump rate ensures that the resulting sequences capture cross-neighborhood relationships rather than simply recapitulating the embedding structure.
    \item \textbf{QA-driven sequences:} For each synthetically generated question-answer pair, we identify the source document and find its embedding neighbors. This simulates a user's browsing trajectory starting from a specific question.
\end{enumerate}

All sequences are augmented $3\times$ through reversal and contiguous subsequence extraction (length 2 to 4), following the augmentation strategy of~\citet{barkan2017item2vec}.

\paragraph{Step 2: Co-occurrence Embedding.} We treat document IDs as tokens and co-occurrence sequences as sentences, training a Word2Vec CBOW model~\citep{mikolov2013efficient}:
\begin{equation}
    \mathbf{z}_d = \text{W2V}_{\text{CBOW}}(d;\; w{=}2,\; n_s{=}10) \quad \mathbf{z}_d \in \mathbb{R}^{100}
\label{eq:w2v}
\end{equation}
The resulting embedding $\mathbf{z}_d$ captures \emph{functional relatedness}: documents that co-occur in user sessions are close in $\mathbf{z}$-space, regardless of their semantic similarity. This is analogous to how Item2Vec~\citep{barkan2017item2vec} learns product representations from co-purchase patterns, except that our ``items'' are knowledge articles and our ``purchases'' are co-access events.

\paragraph{Step 3: Hierarchical Clustering.} We apply agglomerative clustering with cosine distance and average linkage to the co-occurrence embeddings, targeting approximately 20\% of the original KB size. Formally, we compute $\mathcal{C} = \{C_1, \ldots, C_M\}$ where $M{=}\lceil N/5 \rceil$ via agglomerative clustering over $\{\mathbf{z}_d\}_{d \in \mathcal{D}}$. Each cluster $C_j \subset \mathcal{D}$ groups functionally related documents that may span different semantic topics.

\subsection{Online: Hybrid Retrieval with Cluster Expansion}

Given a query $q$ with embedding $\mathbf{e}_q$ produced by any encoder and document embeddings $\{\mathbf{e}_d\}_{d \in \mathcal{D}}$:

\paragraph{Step 1: Vanilla Retrieval.} Retrieve the top-$k_v$ documents by cosine similarity:
\begin{equation}
    \mathcal{V}_q = \underset{d \in \mathcal{D}}{\text{top-}k_v}\; \cos(\mathbf{e}_q, \mathbf{e}_d)
\end{equation}

\paragraph{Step 2: Cluster Expansion.} For each retrieved document, gather all documents from its cluster as expansion candidates:
\begin{equation}
    \mathcal{E}_q = \bigcup_{d \in \mathcal{V}_q} C(d) \setminus \mathcal{V}_q
\end{equation}
where $C(d)$ denotes the cluster containing document $d$. With an average cluster size of 5.0, retrieving $k_v{=}3$ documents expands the candidate pool to approximately $3 \times 5 = 15$ candidates before deduplication.

\paragraph{Step 3: Re-rank and Return.} Rank the expanded candidate set by direct query-document similarity and return the top-$k$:
\begin{equation}
    \mathcal{R}_q = \underset{d \in \mathcal{V}_q \cup \mathcal{E}_q}{\text{top-}k}\; \cos(\mathbf{e}_q, \mathbf{e}_d)
\label{eq:rerank}
\end{equation}

The vanilla results $\mathcal{V}_q$ are guaranteed to appear in the candidate pool, preserving first-hit precision. The expansion adds co-occurring documents that would otherwise be missed because they have low semantic similarity to $q$. We set $k_v{=}3$ and $k{=}8$ throughout our experiments unless otherwise noted.

\subsection{Implementation Details}

For QA pair generation, we use GPT-4.1-nano to produce 10 question-answer pairs per document, yielding 31,500 pairs across the WixQA KB (cost: approximately \$2 USD). Co-occurrence sequence generation produces 16,594 augmented sequences. Word2Vec training completes in under 30 seconds on a single CPU core. Agglomerative clustering over 6,221 documents takes approximately 2 minutes. The entire offline pipeline runs in under 20 minutes, making it practical for periodic KB reorganization.

At inference time, the hybrid retrieval adds negligible latency: cluster lookup is $O(1)$ via a precomputed dictionary, and the expansion step adds at most $k_v \times \bar{|C|}$ candidates (approximately 15) to the re-ranking pool. The total retrieval overhead is dominated by the original embedding similarity computation, not the expansion.

\section{Experimental Setup}
\label{sec:setup}

\subsection{Datasets}

\paragraph{WixQA.} Our primary evaluation uses WixQA~\citep{cohen2025wixqa}, an enterprise RAG benchmark containing 6,221 customer support articles from the Wix Help Center knowledge base. The dataset includes 400 expert-labeled queries: 200 expert-written queries with multi-step answers and 200 simulated queries validated by domain experts. Each query is annotated with ground-truth article IDs from the KB, enabling evaluation with real relevance labels.

\paragraph{E-Commerce Support.} For cross-domain validation, we evaluate on an e-commerce customer support corpus containing 1,000 conversations across 10 product categories and multiple issue types. Co-occurrence sessions are defined by shared issue-category and product-category metadata, providing naturally defined user sessions without synthetic construction.

\subsection{Session Construction}

Sessions correspond to co-occurrence clusters, where each cluster represents a group of articles that collectively serve a user's information-seeking episode. We construct 200 evaluation sessions from the WixQA clusters, with a mean size of 7.9 documents (range: 3 to 70). For each evaluation trial, we randomly sample one entry query from the session and measure retrieval coverage over the full session.

\subsection{Baselines}

We compare our hybrid retrieval against two baselines: (1) \textbf{Vanilla RAG}, which retrieves the top-$k$ documents by cosine similarity between query and document embeddings; and (2) \textbf{Cross-encoder re-ranking}, which retrieves the top-30 by embedding similarity and then re-ranks using ms-marco-MiniLM-L-6-v2 cross-encoder.

\subsection{Embedding Models}

To demonstrate encoder-agnostic gains, we evaluate across four embedding models spanning different architectures, training objectives, and model sizes: all-MiniLM-L6-v2 (22M parameters, general-purpose), bge-base-en-v1.5 (109M, retrieval-optimized), gte-base (109M, text clustering), and e5-base-v2 (109M, weakly supervised).

\subsection{Statistical Methodology}

All reported confidence intervals are computed via bootstrap resampling with 1,000 samples at the 95\% level. We seed all random operations (session entry point selection, sequence generation) with a fixed seed for reproducibility.

\section{Results}
\label{sec:results}

\subsection{Session Coverage vs.\ Retrieval Budget}

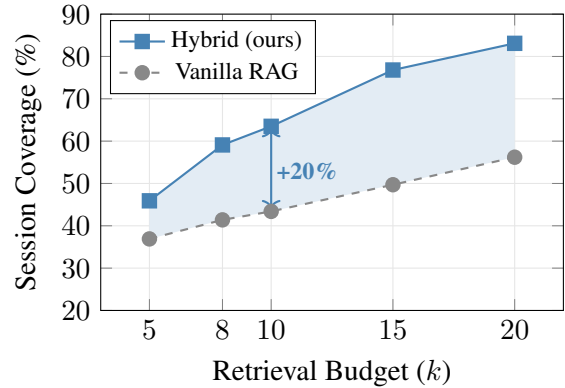
\begin{figure}[t]
\centering
\begin{tikzpicture}
\begin{axis}[
    width=\columnwidth, height=5.5cm,
    xlabel={Retrieval Budget ($k$)},
    ylabel={Session Coverage (\%)},
    xmin=3, xmax=22, ymin=20, ymax=90,
    xtick={5,8,10,15,20},
    ytick={20,30,40,50,60,70,80,90},
    grid=major, grid style={gray!20},
    legend style={at={(0.02,0.98)}, anchor=north west, font=\small},
    every axis plot/.append style={thick},
]
\addplot[name path=upper, hybridcolor, mark=square*, mark size=2.5pt] coordinates {(5,45.9) (8,59.1) (10,63.5) (15,76.8) (20,83.1)};
\addplot[name path=lower, vanillacolor, dashed, mark=*, mark size=2.5pt, mark options={solid}] coordinates {(5,36.9) (8,41.4) (10,43.4) (15,49.7) (20,56.2)};
\addplot[hybridcolor!15] fill between[of=upper and lower];
\legend{Hybrid (ours), Vanilla RAG}
\node[font=\small\bfseries, hybridcolor] at (axis cs:11.5,53) {+20\%};
\draw[hybridcolor, thick, <->] (axis cs:10,44) -- (axis cs:10,63);
\end{axis}
\end{tikzpicture}
\caption{Session coverage vs.\ retrieval budget. The shaded area represents the coverage gap that cluster expansion closes. Notably, the gap \emph{widens} with $k$.}
\label{fig:coverage_budget}
\end{figure}

Table~\ref{tab:coverage_budget} and Figure~\ref{fig:coverage_budget} report session coverage as the retrieval budget varies. The coverage gap between vanilla and hybrid \emph{widens} with $k$: from +9.0\% at $k{=}5$ to +27.1\% at $k{=}15$. This scaling property is important because it means that as context windows grow and retrieval budgets increase, the relative value of cluster expansion grows rather than saturates. At $k{=}20$, hybrid retrieval achieves 83.1\% session coverage, compared to 56.2\% for vanilla.

\begin{table}[t]
\centering
\small
\begin{tabular}{@{}rcccc@{}}
\toprule
$k$ & Vanilla & Hybrid & $\Delta$ & Rel.\ $\Delta$ \\
\midrule
5  & 36.9\% & 45.9\% & +9.0\%  & +24\% \\
8  & 41.4\% & \textbf{59.1\%} & +17.7\% & +43\% \\
10 & 43.4\% & 63.5\% & +20.1\% & +46\% \\
15 & 49.7\% & 76.8\% & +27.1\% & +55\% \\
20 & 56.2\% & 83.1\% & +26.8\% & +48\% \\
\bottomrule
\end{tabular}
\caption{Session coverage at varying retrieval budgets. Bold indicates the primary operating point ($k{=}8$).}
\label{tab:coverage_budget}
\end{table}

\subsection{Retrieval Efficiency}

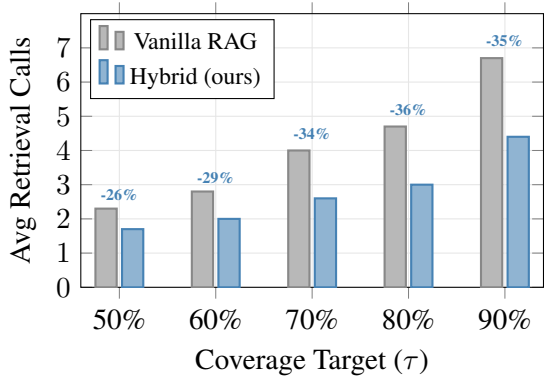
\begin{figure}[t]
\centering
\begin{tikzpicture}
\begin{axis}[
    width=\columnwidth, height=5.2cm,
    ybar,
    bar width=8pt,
    xlabel={Coverage Target ($\tau$)},
    ylabel={Avg Retrieval Calls},
    ymin=0, ymax=8,
    symbolic x coords={50\%,60\%,70\%,80\%,90\%},
    xtick=data,
    ytick={0,1,2,3,4,5,6,7},
    grid=major, grid style={gray!20},
    legend style={at={(0.02,0.98)}, anchor=north west, font=\small},
    every axis plot/.append style={thick},
]
\addplot[fill=vanillacolor!60, draw=vanillacolor] coordinates {(50\%,2.3) (60\%,2.8) (70\%,4.0) (80\%,4.7) (90\%,6.7)};
\addplot[fill=hybridcolor!60, draw=hybridcolor] coordinates {(50\%,1.7) (60\%,2.0) (70\%,2.6) (80\%,3.0) (90\%,4.4)};
\legend{Vanilla RAG, Hybrid (ours)}
\node[font=\tiny\bfseries, hybridcolor] at (axis cs:50\%,2.7) {-26\%};
\node[font=\tiny\bfseries, hybridcolor] at (axis cs:60\%,3.2) {-29\%};
\node[font=\tiny\bfseries, hybridcolor] at (axis cs:70\%,4.5) {-34\%};
\node[font=\tiny\bfseries, hybridcolor] at (axis cs:80\%,5.2) {-36\%};
\node[font=\tiny\bfseries, hybridcolor] at (axis cs:90\%,7.2) {-35\%};
\end{axis}
\end{tikzpicture}
\caption{Retrieval calls to reach session coverage thresholds. Hybrid retrieval requires 26 to 36\% fewer calls across all targets.}
\label{fig:calls}
\end{figure}

Table~\ref{tab:calls} and Figure~\ref{fig:calls} quantify the practical efficiency gain. To reach 70\% session coverage, vanilla RAG requires 4.0 calls on average while hybrid retrieval requires only 2.6, a 34\% reduction. At the 80\% threshold, the savings increase to 1.7 fewer calls per session. In production systems that handle millions of sessions, this translates directly to reduced latency, lower API costs, and fewer user interactions needed to resolve each support episode.

\begin{table}[t]
\centering
\small
\begin{tabular}{@{}rcccc@{}}
\toprule
$\tau$ & Vanilla & Hybrid & Reduction & Saved \\
\midrule
50\% & 2.3 & 1.7 & 26\% & 0.6 \\
60\% & 2.8 & 2.0 & 29\% & 0.8 \\
70\% & 4.0 & \textbf{2.6} & \textbf{34\%} & 1.4 \\
80\% & 4.7 & 3.0 & 36\% & 1.7 \\
90\% & 6.7 & 4.4 & 35\% & 2.4 \\
\bottomrule
\end{tabular}
\caption{Retrieval calls needed to reach coverage thresholds $\tau$.}
\label{tab:calls}
\end{table}

\subsection{Effect of Session Complexity}

Table~\ref{tab:complexity} stratifies results by the number of documents in the session. The method provides substantial gains at every complexity level, with all 95\% bootstrap confidence intervals excluding zero. The largest absolute gain appears on complex sessions (8 to 15 documents): +17.6\% with a CI lower bound of +11.8\%. For simple sessions (3 to 4 documents), coverage jumps from 52.0\% to 70.8\%. The diminishing absolute gain on very complex sessions (16+ documents) is expected: a retrieval budget of $k{=}8$ cannot cover 16 or more documents in a single call regardless of the retrieval strategy.

\begin{table}[t]
\centering
\small
\setlength{\tabcolsep}{4pt}
\begin{tabular}{@{}lrcccc@{}}
\toprule
Type & $N$ & Van. & Hyb. & $\Delta$ & 95\% CI \\
\midrule
Simple (3--4)   & 92 & 52.0 & \textbf{70.8} & +18.8 & [13.1, 24.3] \\
Medium (5--7)   & 48 & 41.5 & 58.1 & +16.7 & [10.7, 22.9] \\
Complex (8--15) & 41 & 28.3 & 45.9 & +17.6 & [11.8, 23.1] \\
V.\ Cmplx (16+) & 19 & 15.2 & 23.8 & +8.6  & [5.5, 11.9] \\
\bottomrule
\end{tabular}
\caption{Session coverage (\%) by complexity with 95\% bootstrap CIs. All intervals exclude zero.}
\label{tab:complexity}
\end{table}

\subsection{Encoder Robustness}

Table~\ref{tab:encoders} demonstrates that the cluster structure, which is learned once from co-occurrence patterns, improves retrieval across all four embedding models. The deltas range from +17.1\% to +20.6\% absolute, with every bootstrap CI excluding zero. This confirms that co-occurrence captures information that is orthogonal to any specific embedding model's representation, and that the offline cluster structure is a genuinely encoder-agnostic resource.

\begin{table}[t]
\centering
\small
\setlength{\tabcolsep}{4pt}
\begin{tabular}{@{}lcccc@{}}
\toprule
Encoder & Van. & Hyb. & $\Delta$ & 95\% CI \\
\midrule
MiniLM-L6-v2  & 41.1 & 58.2 & +17.1 & [14.0, 20.2] \\
bge-base-v1.5 & 39.8 & 57.1 & +17.2 & [14.2, 20.7] \\
gte-base      & 40.4 & 60.8 & +20.4 & [17.0, 23.8] \\
e5-base-v2    & 39.8 & \textbf{60.4} & +20.6 & [17.1, 23.9] \\
\bottomrule
\end{tabular}
\caption{Session coverage (\%) across encoders ($k{=}8$) with 95\% bootstrap CIs. All intervals exclude zero. Clusters are learned once and benefit all encoders.}
\label{tab:encoders}
\end{table}

\subsection{Domain Generalization}

We apply Latent Dirichlet Allocation to identify six functional domains within WixQA (Table~\ref{tab:domains}). The hybrid method improves coverage in \textbf{all six domains}, with gains ranging from +12.7\% (Apps \& Email) to +21.5\% (Editor \& Studio). This consistency across topically diverse domains confirms that the method is not specific to any particular content type or user behavior pattern within the KB.

\begin{table}[t]
\centering
\small
\begin{tabular}{@{}lrccc@{}}
\toprule
Domain & $N$ & Van. & Hyb. & $\Delta$ \\
\midrule
Bookings \& Payments & 37 & 34.2\% & 50.8\% & +16.6\% \\
Apps \& Email         & 28 & 40.0\% & 52.6\% & +12.7\% \\
Editor \& Studio      & 58 & 46.3\% & \textbf{67.8\%} & +21.5\% \\
Plans \& Pricing      & 13 & 52.8\% & 71.1\% & +18.3\% \\
Media \& Domains      & 26 & 50.0\% & 63.0\% & +13.0\% \\
Stores \& Blog/CMS    & 38 & 37.6\% & 51.9\% & +14.3\% \\
\bottomrule
\end{tabular}
\caption{Session coverage across six WixQA functional domains. Gains are positive in all six domains.}
\label{tab:domains}
\end{table}

\subsection{Cross-Dataset Validation}

To validate beyond a single benchmark, we evaluate on an e-commerce customer support corpus with 1,000 documents and 94 naturally defined sessions. Session coverage improves from 32.2\% to 43.3\% (+11.1\% absolute, +34\% relative), confirming that the method generalizes to a different domain, document structure, and session definition.

\section{Analysis}
\label{sec:analysis}

\subsection{Semantic Similarity vs.\ Co-occurrence}

\begin{figure}[t]
\centering
\begin{tikzpicture}
\begin{axis}[
    width=\columnwidth, height=5.5cm,
    xlabel={Semantic Similarity (cosine)},
    ylabel={Co-occurrence Strength},
    xmin=-0.05, xmax=1.05, ymin=-0.5, ymax=8,
    grid=major, grid style={gray!15},
    legend style={at={(0.98,0.02)}, anchor=south east, font=\small},
]
\addplot[only marks, mark=*, mark size=0.8pt, gray!40] coordinates {
(0.12,0.1) (0.25,0.2) (0.38,0.0) (0.52,0.1) (0.68,0.3) (0.82,0.1) (0.15,0.4)
(0.33,0.2) (0.47,0.0) (0.61,0.3) (0.75,0.1) (0.88,0.2) (0.05,0.1) (0.92,0.0)
(0.21,0.3) (0.44,0.1) (0.56,0.2) (0.71,0.0) (0.35,0.4) (0.58,0.1) (0.78,0.3)
(0.42,0.2) (0.65,0.1) (0.85,0.4) (0.18,0.0) (0.28,0.3) (0.49,0.2) (0.73,0.1)
};
\addplot[only marks, mark=square*, mark size=2pt, accentgreen!80!black] coordinates {
(0.72,5.2) (0.81,6.1) (0.76,4.8) (0.85,5.5) (0.69,4.2) (0.78,6.8) (0.83,5.9)
(0.74,7.1) (0.88,4.5) (0.71,5.8) (0.79,6.3) (0.86,4.9) (0.73,5.1) (0.82,6.5)
};
\addplot[only marks, mark=triangle*, mark size=2.5pt, accentred] coordinates {
(0.08,5.5) (0.15,6.2) (0.22,4.8) (0.11,7.1) (0.28,5.1) (0.18,6.5) (0.05,4.3)
(0.25,5.8) (0.13,6.9) (0.31,4.5) (0.09,5.3) (0.21,7.5) (0.16,5.0) (0.27,6.1)
(0.19,4.7) (0.07,6.8) (0.24,5.4) (0.14,7.2)
};
\legend{Non-co-occurring, High-sim high-cooc, \textbf{Low-sim high-cooc}}
\draw[accentred, thick, dashed, rounded corners=2pt] (axis cs:-0.02,3.8) rectangle (axis cs:0.35,7.8);
\node[font=\scriptsize\bfseries, accentred, text width=2cm, align=center] at (axis cs:0.17,3.2) {Vanilla RAG\\misses these};
\node[font=\scriptsize, fill=white, draw=gray!40, rounded corners=2pt, inner sep=2pt] at (axis cs:0.7,1.5) {$r = 0.20$};
\end{axis}
\end{tikzpicture}
\caption{Semantic similarity vs.\ co-occurrence strength (schematic from 10K sampled pairs). The \textcolor{accentred}{\textbf{red triangle}} region (low similarity, high co-occurrence) contains document pairs that vanilla RAG cannot surface together. Weak Pearson correlation ($r{=}0.20$) confirms the signals are complementary.}
\label{fig:gap}
\end{figure}
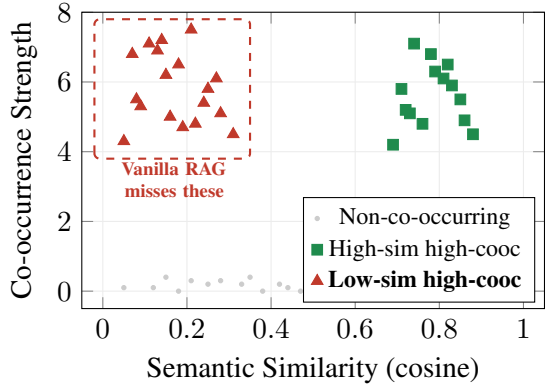

We compute the Pearson correlation between semantic similarity (cosine in MiniLM embedding space) and co-occurrence strength for 10,000 sampled document pairs (Figure~\ref{fig:gap}). The correlation is weak ($r \approx 0.2$), confirming that documents users need together are \emph{not} necessarily the documents that look alike in embedding space. This weak correlation is precisely what makes co-occurrence clustering valuable: it provides a complementary signal that semantic embeddings structurally cannot capture, and cluster expansion surfaces the ``low similarity, high co-occurrence'' document pairs that vanilla RAG systematically misses.

\subsection{Cluster Quality Analysis}

Table~\ref{tab:cluster_stats} reports detailed clustering statistics. The 1,244 clusters have a mean size of 5.0 documents with a long-tailed distribution: most clusters contain 3 to 5 documents (corresponding to simple user sessions), while a small number of large clusters (up to 70 documents) capture broad topics like ``account management'' that touch many articles.

\begin{table}[t]
\centering
\small
\begin{tabular}{@{}lr@{}}
\toprule
Statistic & Value \\
\midrule
Original KB size & 6,221 documents \\
Number of clusters $M$ & 1,244 \\
Compression ratio & 20.0\% \\
Mean cluster size & 5.0 \\
Median cluster size & 4 \\
Max cluster size & 70 \\
Silhouette score & 0.205 \\
Within-cluster co-occ rate & 25.9\% \\
Word2Vec vocabulary coverage & 98.8\% \\
Augmented co-occ sequences & 16,594 \\
\bottomrule
\end{tabular}
\caption{Clustering statistics on WixQA.}
\label{tab:cluster_stats}
\end{table}

The silhouette score of 0.205 indicates moderate cluster cohesion. This moderate (rather than high) score is expected and actually desirable: co-occurrence clusters intentionally group semantically \emph{diverse} documents, so high cohesion in the semantic embedding space would indicate that the clusters are simply recapitulating semantic similarity rather than adding new information.

The within-cluster co-occurrence rate of 25.9\% means that approximately one in four co-occurring document pairs land in the same cluster. This is substantially above the random baseline of $1/M \approx 0.08\%$, confirming meaningful structure. The rate is not higher because the co-occurrence sequences draw from three signal sources with different properties, and the clustering finds a compromise that respects all three.

\subsection{Qualitative Example}

To illustrate the method's behavior concretely, consider a representative cluster:

\begin{table}[t]
\centering
\small
\begin{tabular}{@{}cl@{}}
\toprule
\# & Document Title \\
\midrule
1 & Connecting a Domain You Already Own \\
2 & Transferring a Domain to Wix \\
3 & Getting a Free Domain with Premium Plans \\
4 & Changing Your Site Template \\
5 & Customizing Your Site's Favicon \\
\bottomrule
\end{tabular}
\caption{Example cluster containing semantically diverse but functionally related documents. Documents 1--3 cover domain management; documents 4--5 cover site design. Users setting up a new website commonly need all five.}
\label{tab:qualitative}
\end{table}

Documents 1 through 3 form a tight semantic group (domain management), while documents 4 and 5 are semantically distant (site design). Yet all five are commonly needed by users setting up a new website. When a user asks about connecting their domain, vanilla RAG retrieves documents 1 through 3 but misses 4 and 5 entirely. Cluster expansion adds documents 4 and 5 to the candidate pool, and the re-ranker surfaces them if they match the query context.

\subsection{Precision Trade-off}

Cluster expansion trades marginal first-hit precision for substantially broader coverage. At $k{=}8$, Hits@K decreases slightly from 96.0\% to 93.0\%. This is the correct trade-off for session-oriented settings: covering 58\% of the user's full information need is more valuable than a 3\% improvement in whether the single best document appears in the top result. In practice, this trade-off can be tuned by adjusting $k_v$: increasing $k_v$ from 3 to 5 preserves more vanilla precision at the cost of fewer expansion slots.

\subsection{Boundary Condition: When the Method Does Not Help}

On HotpotQA~\citep{yang2018hotpotqa}, where each question has an isolated paragraph set with no cross-query document reuse, the method shows no gain ($\Delta = -1.4\%$). This negative result is informative: it confirms that the improvement requires a \emph{persistent} knowledge base where documents are reused across multiple user sessions. In such settings, co-occurrence patterns form naturally from repeated access. In benchmarks where each question constructs an ad-hoc document set, no co-occurrence signal can emerge. This boundary condition confirms that gains are driven by learned co-occurrence patterns rather than by an artifact of the expansion mechanism itself.

\section{Conclusion}

We propose that RAG evaluation should shift from single-query recall to session-level coverage, and demonstrate a simple method to close the resulting coverage gap: reorganize the KB offline using co-occurrence-aware clustering, then expand retrieval candidates through cluster neighborhoods at query time. A single retrieval call over a reorganized KB covers 58\% of a user's session-level information need (compared to 41\% for standard RAG) while requiring 34\% fewer calls to reach practical coverage thresholds. The method is encoder-agnostic (+17 to 21\% across four models), domain-general (gains in 6 out of 6 domains), and provides 80\% KB compression as a side benefit.

For future work, we plan to incorporate real user navigation logs to replace synthetic co-occurrence sequences and measure how the quality of the co-occurrence signal affects downstream coverage gains. We also plan to evaluate the impact on end-to-end generation quality using LLM-as-judge metrics, as higher retrieval coverage does not automatically translate to better generated answers if the LLM cannot effectively use the broader context. Finally, we intend to explore adaptive cluster expansion, where the number of expanded candidates scales with query ambiguity, and to investigate combining our pre-retrieval KB reorganization with post-retrieval compression methods like EDC$^2$-RAG~\citep{li2025efficient} for end-to-end session-level RAG optimization.

\section*{Limitations}

Our co-occurrence sequences are constructed from embedding neighborhoods and synthetic QA pairs rather than real user interaction logs. While the ground-truth article groups from WixQA provide a real co-occurrence signal and the HotpotQA boundary experiment confirms that persistent document reuse is required, real enterprise usage data would likely produce stronger co-occurrence patterns and larger gains. We were unable to use real navigation logs due to the absence of publicly available enterprise session data, a common constraint in this domain.

Session-level evaluation uses cluster-derived sessions as a proxy for real user sessions; evaluation on logged multi-turn conversations would strengthen the claims.

We evaluate retrieval coverage but not downstream generation quality. Higher coverage introduces more diverse context into the LLM's input, which may trigger the ``lost in the middle'' phenomenon~\citep{liu2024lost} where models underutilize information in the middle of long contexts. Whether the 58\% session coverage translates to proportionally better generated answers remains an open question that we leave to future work.

The 40\% random jump probability in embedding walks was set heuristically; a systematic search over this hyperparameter could improve results. Finally, while we evaluate across four encoders, six domains, and two datasets, additional enterprise KB evaluations with real user session logs would further establish generalizability.

\section*{Ethics Statement}

This work uses publicly available datasets (WixQA under MIT license) and does not involve human subjects, private user data, or personally identifiable information. The synthetic QA pairs are generated from public knowledge base articles. Our method reorganizes existing knowledge bases and does not generate new content, so it does not introduce additional hallucination risk beyond what is inherent in the underlying RAG system. We note that co-occurrence patterns learned from real user logs, if used in future work, would require appropriate anonymization and privacy protections.

\bibliography{custom}

@misc{lewis2021retrievalaugmented,
      title={Retrieval-Augmented Generation for Knowledge-Intensive {NLP} Tasks}, 
      author={Patrick Lewis and Ethan Perez and Aleksandra Piktus and Fabio Petroni and Vladimir Karpukhin and Naman Goyal and Heinrich K{\"u}ttler and Mike Lewis and Wen-tau Yih and Tim Rockt{\"a}schel and Sebastian Riedel and Douwe Kiela},
      year={2021},
      eprint={2005.11401},
      archivePrefix={arXiv},
      primaryClass={cs.CL},
      url={https://arxiv.org/abs/2005.11401}, 
}

@misc{mikolov2013efficient,
      title={Efficient Estimation of Word Representations in Vector Space}, 
      author={Tomas Mikolov and Kai Chen and Greg Corrado and Jeffrey Dean},
      year={2013},
      eprint={1301.3781},
      archivePrefix={arXiv},
      primaryClass={cs.CL},
      url={https://arxiv.org/abs/1301.3781}, 
}

@misc{cohen2025wixqa,
      title={{WixQA}: A Multi-Dataset Benchmark for Enterprise Retrieval-Augmented Generation}, 
      author={Dvir Cohen and Lin Burg and Sviatoslav Pykhnivskyi and Hagit Gur and Stanislav Kovynov and Olga Atzmon and Gilad Barkan},
      year={2025},
      eprint={2505.08643},
      archivePrefix={arXiv},
      primaryClass={cs.AI},
      url={https://arxiv.org/abs/2505.08643}, 
}

@inproceedings{li2025efficient,
    title={Efficient Dynamic Clustering-Based Document Compression for Retrieval-Augmented-Generation},
    author={Li, Weitao and Zhang, Xiangyu and Liu, Kaiming and Lei, Xuanyu and Ma, Weizhi and Liu, Yang},
    booktitle={Findings of the Association for Computational Linguistics: EMNLP 2025},
    month=nov,
    year={2025},
    address={Suzhou, China},
    publisher={Association for Computational Linguistics},
    url={https://aclanthology.org/2025.findings-emnlp.522/},
    doi={10.18653/v1/2025.findings-emnlp.522},
    pages={9833--9849},
}

@misc{xu2025clusterbased,
      title={Cluster-based Adaptive Retrieval: Dynamic Context Selection for {RAG} Applications}, 
      author={Yifan Xu and Vipul Gupta and Rohit Aggarwal and Varsha Mahadevan and Bhaskar Krishnamachari},
      year={2025},
      eprint={2511.14769},
      archivePrefix={arXiv},
      primaryClass={cs.IR},
      url={https://arxiv.org/abs/2511.14769}, 
}

@misc{akesson2024crag,
      title={Clustered Retrieved Augmented Generation ({CRAG})}, 
      author={Simon Akesson and Frances A. Santos},
      year={2024},
      eprint={2406.00029},
      archivePrefix={arXiv},
      primaryClass={cs.CL},
      url={https://arxiv.org/abs/2406.00029}, 
}

@inproceedings{es2024ragas,
    title={{RAGAS}: Automated Evaluation of Retrieval Augmented Generation},
    author={Es, Shahul and James, Jithin and Espinosa Anke, Luis and Schockaert, Steven},
    booktitle={Proceedings of the 18th Conference of the European Chapter of the Association for Computational Linguistics: System Demonstrations},
    month=mar,
    year={2024},
    address={St. Julians, Malta},
    publisher={Association for Computational Linguistics},
    url={https://aclanthology.org/2024.eacl-demo.16/},
    doi={10.18653/v1/2024.eacl-demo.16},
    pages={150--158},
}

@inproceedings{yang2018hotpotqa,
    title={{H}otpot{QA}: A Dataset for Diverse, Explainable Multi-hop Question Answering},
    author={Yang, Zhilin and Qi, Peng and Zhang, Saizheng and Bengio, Yoshua and Cohen, William and Salakhutdinov, Ruslan and Manning, Christopher D.},
    booktitle={Proceedings of the 2018 Conference on Empirical Methods in Natural Language Processing},
    month=oct,
    year={2018},
    address={Brussels, Belgium},
    publisher={Association for Computational Linguistics},
    url={https://aclanthology.org/D18-1259/},
    doi={10.18653/v1/D18-1259},
    pages={2369--2380},
}

@misc{ma2023finetuning,
      title={Fine-Tuning {LLaMA} for Multi-Stage Text Retrieval}, 
      author={Xueguang Ma and Liang Wang and Nan Yang and Furu Wei and Jimmy Lin},
      year={2023},
      eprint={2310.08319},
      archivePrefix={arXiv},
      primaryClass={cs.IR},
      url={https://arxiv.org/abs/2310.08319}, 
}

@misc{weller2026theoretical,
      title={On the Theoretical Limitations of Embedding-Based Retrieval}, 
      author={Orion Weller and Michael Boratko and Iftekhar Naim and Jinhyuk Lee},
      year={2026},
      eprint={2508.21038},
      archivePrefix={arXiv},
      primaryClass={cs.IR},
      url={https://arxiv.org/abs/2508.21038}, 
}

@misc{barkan2017item2vec,
      title={Item2Vec: Neural Item Embedding for Collaborative Filtering}, 
      author={Oren Barkan and Noam Koenigstein},
      year={2017},
      eprint={1603.04259},
      archivePrefix={arXiv},
      primaryClass={cs.LG},
      url={https://arxiv.org/abs/1603.04259}, 
}

@article{liu2024lost,
    title={Lost in the Middle: How Language Models Use Long Contexts},
    author={Liu, Nelson F. and Lin, Kevin and Hewitt, John and Paranjape, Ashwin and Bevilacqua, Michele and Petroni, Fabio and Liang, Percy},
    journal={Transactions of the Association for Computational Linguistics},
    volume={12},
    pages={157--173},
    year={2024},
    doi={10.1162/tacl_a_00638},
}

@misc{sarthi2024raptor,
      title={{RAPTOR}: Recursive Abstractive Processing for Tree-Organized Retrieval}, 
      author={Parth Sarthi and Salman Abdullah and Aditi Tuli and Shubh Khanna and Anna Goldie and Christopher D. Manning},
      year={2024},
      eprint={2401.18059},
      archivePrefix={arXiv},
      primaryClass={cs.CL},
      url={https://arxiv.org/abs/2401.18059}, 
}

@inproceedings{gao2023precise,
    title={Precise Zero-Shot Dense Retrieval without Relevance Labels},
    author={Gao, Luyu and Ma, Xueguang and Lin, Jimmy and Callan, Jamie},
    booktitle={Proceedings of the 61st Annual Meeting of the Association for Computational Linguistics (Volume 1: Long Papers)},
    month=jul,
    year={2023},
    address={Toronto, Canada},
    publisher={Association for Computational Linguistics},
    url={https://aclanthology.org/2023.acl-long.99/},
    doi={10.18653/v1/2023.acl-long.99},
    pages={1762--1777},
}

\appendix

\section{Single-Query Retrieval Results}
\label{app:singlequery}

For completeness, we report standard single-query retrieval metrics on the 400 WixQA expert queries (Table~\ref{tab:singlequery}). We evaluate a boost-based scoring variant which adds a cluster-relevance signal to direct document similarity with weight $\alpha{=}0.2$. This provides modest but consistent gains over vanilla retrieval on single-query metrics. The gains are small because WixQA queries are predominantly single-aspect; the method's primary value is at the session level (Section~\ref{sec:results}).

\begin{table}[h]
\centering
\small
\setlength{\tabcolsep}{4pt}
\begin{tabular}{@{}lccc@{}}
\toprule
Method & W.\ Recall & Hits@K & MRR \\
\midrule
Vanilla       & 52.8\% & 57.5\% & 0.359 \\
Reranker      & 53.1\% & 58.0\% & 0.334 \\
Boost ($\alpha{=}0.2$) & 55.2\% & 59.8\% & 0.363 \\
\bottomrule
\end{tabular}
\caption{Single-query metrics on 400 WixQA expert queries.}
\label{tab:singlequery}
\end{table}

\section{Hyperparameter Sensitivity}
\label{app:hyperparams}

\paragraph{Vanilla slots $k_v$.} Table~\ref{tab:kv_sweep} shows session coverage as $k_v$ varies with $k{=}8$ fixed. Lower $k_v$ allocates more slots to cluster expansion, increasing coverage at the cost of first-hit precision. We select $k_v{=}3$ as it provides the best balance.

\begin{table}[h]
\centering
\small
\begin{tabular}{@{}rccc@{}}
\toprule
$k_v$ & Coverage & Hits@K & $\Delta$ Cov \\
\midrule
1 & 61.2\% & 87.5\% & +20.1\% \\
2 & 59.8\% & 90.0\% & +18.7\% \\
3 & 58.2\% & 93.0\% & +17.1\% \\
5 & 50.2\% & 95.5\% & +9.1\% \\
7 & 43.5\% & 96.0\% & +2.4\% \\
\bottomrule
\end{tabular}
\caption{Effect of vanilla slots $k_v$ on session coverage and Hits@K ($k{=}8$). $k_v{=}3$ balances coverage with precision.}
\label{tab:kv_sweep}
\end{table}

\paragraph{Boost weight $\alpha$.} For the boost-based variant used in single-query evaluation, Table~\ref{tab:alpha_sweep} shows the effect of $\alpha$. The value $\alpha{=}0.2$ provides the best weighted recall; higher values degrade performance as the cluster signal overwhelms direct similarity.

\begin{table}[h]
\centering
\small
\begin{tabular}{@{}rccc@{}}
\toprule
$\alpha$ & W.\ Recall & Hits@K & $\Delta$ WR \\
\midrule
0.1 & 54.1\% & 58.8\% & +1.2\% \\
0.2 & \textbf{55.2\%} & 59.8\% & +2.4\% \\
0.3 & 54.2\% & 58.8\% & +1.3\% \\
0.4 & 52.2\% & 56.5\% & $-$0.7\% \\
0.5 & 50.8\% & 55.0\% & $-$2.0\% \\
\bottomrule
\end{tabular}
\caption{Effect of boost weight $\alpha$ on single-query metrics.}
\label{tab:alpha_sweep}
\end{table}

\section{Cross-Dataset Summary}
\label{app:crossdataset}

Table~\ref{tab:crossdataset_full} summarizes results across all evaluation settings, including the HotpotQA boundary experiment.

\begin{table}[h]
\centering
\small
\setlength{\tabcolsep}{3pt}
\begin{tabular}{@{}llrccr@{}}
\toprule
Dataset & Domain & Docs & Van. & Hyb. & $\Delta$ \\
\midrule
WixQA & Web Support & 6,221 & 41\% & 58\% & +17\% \\
E-Commerce & E-Commerce & 1,000 & 32\% & 43\% & +11\% \\
\midrule
HotpotQA & Open-domain & 5,842 & 35.8\% & 34.4\% & $-$1.4\% \\
\bottomrule
\end{tabular}
\caption{Cross-dataset session coverage ($k{=}8$). Gains are consistent on persistent KBs and absent on ad-hoc paragraph sets (HotpotQA), confirming the method requires cross-session document reuse.}
\label{tab:crossdataset_full}
\end{table}

\section{Reproducibility Details}
\label{app:reproducibility}

\begin{table}[h]
\centering
\small
\begin{tabular}{@{}ll@{}}
\toprule
Parameter & Value \\
\midrule
Word2Vec architecture & CBOW \\
Embedding dimension & 100 \\
Window size & 2 \\
Negative samples & 10 \\
Training epochs & 30 \\
Random jump probability & 0.40 \\
Clustering method & Agglomerative (cosine, avg) \\
Target compression & 20\% ($M = \lceil N/5 \rceil$) \\
Vanilla slots $k_v$ & 3 \\
Retrieval budget $k$ & 8 \\
QA generation model & GPT-4.1-nano \\
QA pairs per document & 10 \\
Bootstrap samples & 1,000 \\
Random seed & 42 \\
\bottomrule
\end{tabular}
\caption{Full hyperparameter configuration.}
\label{tab:reproducibility}
\end{table}

The offline pipeline (QA generation, sequence construction, Word2Vec training, and clustering) runs in under 20 minutes on a single CPU core, excluding API time for QA generation (approximately 6 minutes with 30 concurrent threads). All embedding models are loaded via the \texttt{sentence-transformers} library. WixQA is available under MIT license at \texttt{Wix/WixQA} on HuggingFace.

\end{document}